\documentclass[twocolumn,twoside]{revtex4}
\usepackage{graphicx}
\usepackage{fancyhdr}
\begin{document}

\title{The search of axion-like-particles with Fermi and Cherenkov telescopes\footnote{A summary of the work presented in Ref.\cite{masc_axions}}}

%

\author{Miguel A. S\'anchez-Conde}
\affiliation{Instituto de Astrof\'isica de Canarias, E-38205 La Laguna, Tenerife, Spain}
\author{David Paneque}
\affiliation{Kavli Institute for Particle Astrophysics and Cosmology (KIPAC), SLAC National Accelerator Center, Sand Hill Road 2575, CA 94025, USA}

\author{Elliott Bloom}
\affiliation{Kavli Institute for Particle Astrophysics and Cosmology (KIPAC), SLAC National Accelerator Center, Sand Hill Road 2575, CA 94025, USA}

\author{Francisco Prada}
\affiliation{Instituto de Astrof\'isica de Andaluc\'ia (CSIC), E-18008, Granada, Spain}

\author{Alberto Dom\'inguez}
\affiliation{Instituto de Astrof\'isica de Andaluc\'ia (CSIC), E-18008, Granada, Spain}
\affiliation{Departamento de F\'isica At\'omica, Molecular y Nuclear, Universidad de Sevilla, E-41012, Sevilla, Spain}

\begin{abstract}
Axion Like Particles (ALPs), postulated to solve the strong-CP problem, are predicted to couple with photons in the presence of magnetic fields, which may lead to a significant change in the observed spectra of gamma-ray sources such as AGNs. Here we simultaneously consider in the same framework both the photon/axion mixing that takes place in the gamma-ray source and that one expected to occur in the intergalactic magnetic fields. We show that photon/axion mixing could explain recent puzzles regarding the observed spectra of distant gamma-ray sources as well as the recently published lower limit to the EBL intensity. We finally summarize the different signatures expected and discuss the best strategy to search for ALPs with the Fermi satellite and current Cherenkov telescopes like CANGAROO, HESS, MAGIC and VERITAS.

\end{abstract}

\maketitle

\thispagestyle{fancy}

\section{Photon/axion oscillations}    \label{sec1}

Axions were postulated to solve the strong-CP problem in QCD in the 1970s \cite{PQ}, and are still the most compelling explanation for it. Moreover, they are also appealing for astrophysicists, since they are valid dark matter candidates to constitute a portion or the totality of the non-barionic cold dark matter content predicted to exist in the Universe.  But probably the most interesting property of axions, or in a more generic way, Axion-Like Particles (ALPs), where the mass $m_a$ and the coupling constant are not related to each other, is that they are expected to oscillate into photons (and viceversa) in the presence of magnetic fields \cite{dicus,sikivie}. This oscillation of photons to ALPs are the main vehicle used at present in axion searches, like those carried out by CAST \cite{cast} or ADMX \cite{admx}, but they could also have important implications for astronomical observations. For example, they could distort the spectra of gamma-ray sources, such as Active Galactic Nuclei (AGNs) \cite{hooper,deangelis,hochmuth,simet} or galactic sources in the TeV range \cite{mirizzi07}. These distortions may be detected by current gamma-ray experiments, such as Imaging Atmospheric Cherenkov Telescopes (IACTs) like MAGIC \cite{magic}, HESS \cite{hess}, VERITAS \cite{veritas} or CANGAROO-III \cite{cangaroo}, covering energies in the range 0.1-20 TeV, and by the Fermi satellite \cite{glast}, which operates at energies in the range 0.02-300 GeV.

Given a domain of length $s$, where there is a roughly constant magnetic field and plasma frequency, the probability of a photon of energy $E_{\gamma}$ to be converted into an ALP after traveling through it can be written as \cite{mirizzi07,hochmuth}:

\begin{equation}
P_0 = (\Delta_B~s)^2~\frac{\sin^2(\Delta_{osc}~s/2)}{(\Delta_{osc}~s/2)^2} 
\label{eq:P0}
\end{equation}

\noindent Here $\Delta_{osc}$ is the oscillation wave number: 
\begin{equation}
\Delta_{osc}^2 \simeq (\Delta_{CM}+\Delta_{pl}-\Delta_a)^2+4\Delta_B^2,
\label{eq:deltaosc}
\end{equation}

\noindent $\Delta_B$ that gives us an idea of how effective is the mixing, i.e.
\begin{equation}
\Delta_B = \frac{B_t}{2~M} \simeq 1.7 \times 10^{-21}~M_{11}~B_{mG}~cm^{-1},
\label{eq:deltaB}
\end{equation}
\noindent where $B_t$ the magnetic field component along the polarization vector of the photon and $M_{11}$ the inverse of the coupling constant.

$\Delta_{CM}$ is the vacuum Cotton-Mouton term, i.e.
\begin{eqnarray}
\Delta_{CM} &=& -\frac{\alpha}{45\pi}~\left(\frac{B_t}{B_{cr}}\right)^2E_{\gamma} \nonumber \\
&\simeq& -1.3 \times 10^{-21}~B^2_{mG}\left(\frac{E_{\gamma}}{TeV}\right)~cm^{-1}, 
\label{eq:deltaCM}
\end{eqnarray}
\noindent where $B_{cr}=m^2_e/e \simeq 4.41 \times 10^{13}$~G the critical magnetic field strength ($e$ is the electron charge).

$\Delta_{pl}$ is the plasma term:
\begin{equation}
\Delta_{pl} = \frac{w^2_{pl}}{2E} \simeq 3.5 \times 10^{-20}\left(\frac{n_e}{10^3cm^{-3}}\right)\left(\frac{TeV}{E_{\gamma}}\right)~cm^{-1},
\label{eq:deltapl}
\end{equation}
\noindent where $w_{pl}=\sqrt{4\pi\alpha n_e/m_e} = 0.37 \times 10^{-4} \mu eV \sqrt{n_e/cm^{-3}}$ the plasma frequency, $m_e$ the electron mass and $n_e$ the electron density.

Finally, $\Delta_a$ is the ALP mass term:
\begin{equation}
\Delta_{a} = \frac{m^2_{a}}{2E_{\gamma}} \simeq 2.5 \times 10^{-20}m^2_{a,\mu eV}\left(\frac{TeV}{E_{\gamma}}\right)~cm^{-1}.
\label{eq:deltaa}
\end{equation}

Note that in Eqs.(\ref{eq:deltaB}-\ref{eq:deltaa}) we have introduced the dimensionless quantities $B_{mG}=B/10^{-3}$ G, $M_{11}=M/10^{11}$ GeV and $m_{\mu eV}=m/10^{-6}$ eV.

Since we expect to have not only one coherence domain but several domains with magnetic fields different from zero and subsequently with a potential photon/axion mixing in each of them, we can derive a total conversion probability \cite{mirizzi07} as follows:

\begin{equation}
P_{\gamma \rightarrow a} \simeq \frac{1}{3}[1-\exp(-3NP_0/2)]
\label{eq:totalProb}
\end{equation}

\noindent where $P_0$ is given by Eq.(\ref{eq:P0}) and $N$ represents the number of domains. Note that in the limit where $N~P_0 \rightarrow \infty$, the total probability saturates to 1/3, i.e. one third of the photons will convert into ALPs.

It is useful here to rewrite Eq.~(\ref{eq:P0}) following Ref.~\cite{hooper}, i.e.

\begin{equation}\label{eq:prob2}
P_0 =\frac{1}{1+(E_{crit}/E_{\gamma})^2}~
\sin^2\left[\frac{B~s}{2~M}\sqrt{1+\left(\frac{E_{crit}}{E_{\gamma}}\right)^2}\right]
\end{equation}

\noindent so that we can define a characteristic energy, $E_{crit}$, given by:
\begin{equation} 
E_{crit} \equiv \frac{m^2~M}{2~B}
\label{eq:ecrit1}
\end{equation}

\noindent or in more convenient units:
\begin{equation}
E_{crit} (GeV) \equiv \frac{m^2_{\mu eV}~M_{11}}{0.4~B_G}
\label{eq:ecrit}
\end{equation}

\noindent where the subindices refer again to dimensionless quantities: $m_{\mu eV} \equiv m/ \mu eV$, $M_{11} \equiv M/10^{11}$ GeV and $B_G \equiv $ B/Gauss; $m$ is the effective ALP mass $m^2 \equiv |m_a^2-\omega_{pl}^2|$. Recent results from the CAST experiment \cite{cast} give a value of $M_{11} \geq 0.114$ for axion mass $m_a \leq 0.02$ eV. Although there are other limits derived with other methods or experiments, the CAST bound is the most general and stringent limit in the range $10^{-11}$ eV $\ll m_a \ll 10^{-2}$ eV.

\noindent In order to have an efficient conversion, we need \cite{hooper}: 
\begin{equation} 
\frac{15~B_G~s_{pc}}{M_{11}} \ge 1 
\label{eq:effconv}
\end{equation}
\noindent where s$_{pc} \equiv $ s/parsec. Some astrophysical environments fulfill the above mixing requirement and the M$_{11}$ constraints imposed by CAST. Indeed, when using $M_{11} \geq 0.114$ in Eq.~(\ref{eq:effconv}), we can deduce that astrophysical sources with B$_G \cdot s_{pc} \ge$ 0.01 will be valid.  This product B$ \cdot s$ also determines the maximum energy E$_{max}$ to which sources can accelerate cosmic rays, and is given by E$_{max}=9.3 \times 10^{20}$ B$_G \cdot s_{pc}$ eV (Hillas criterium). Since we observe cosmic rays up to 3 $\times$ 10$^{20}$ eV, B$_G$~s$_{pc}$ can be as high as 0.3, which also means that sources with B$_G \cdot s_{pc} = $ 0.01 are completely plausible and should exist. Therefore, photon/axion mixing may have important implications for astronomical observations (AGNs, pulsars, GRBs...). Note, however, that an efficient mixing can be expected to occur not only in compact sources: the mixing will be also present in the Intergalactic Magnetic Fields (IGMFs), with typical values of $\sim$1 nG for the B field, provided that the source is located at cosmological distances (s$_{pc}=10^8$) in order to ensure that B $\cdot s$ is still larger than 0.01. 

\noindent Therefore, in order to correctly evaluate the photon/axion mixing effect for distant sources, it will be necessary to handle under the same consistent framework the mixing that takes place inside or near the gamma-ray sources together with that one expected to occur in the IGMF. In the literature, both effects have been considered separately. We neglect, however, the mixing that may happen inside the Milky Way due to galactic magnetic fields.\footnote{At present, a concise modeling of this effect is still very dependent on the largely unknown morphology of the magnetic field in the galaxy. Furthermore, in the most optimistic case, it would produce a photon flux enhancement of $\sim$3\% of the initial photon flux emitted by the source \cite{simet}.}

\subsection{Mixing in the source}
\begin{figure}[!h]
\centering
\includegraphics[height=6.5cm,width=8cm]{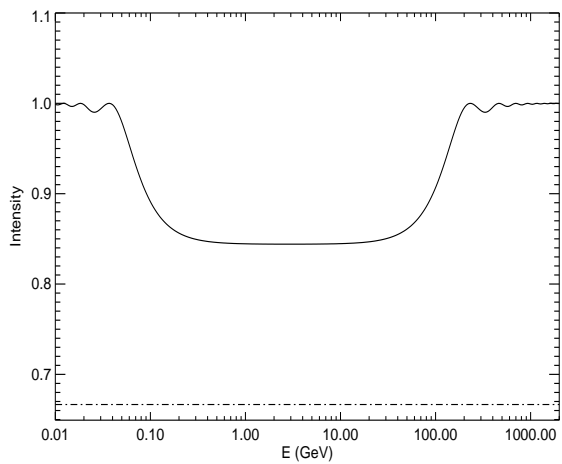}
\caption{\small{Example of photon/axion oscillations inside the source or vicinity, and its effect on the source intensity (solid line), which was normalized to 1 in the Figure.  We used the parameters given in Table~\ref{tab:3c279} to model the AGN source, but we adopted an ALP mass of 1 $\mu$eV. This gives $E_{crit}=0.19$ GeV. The dot-dashed line represents the maximum (theoretical) attenuation, i.e. 1/3.}} 
\label{fig:sourcemix}
\end{figure}

To illustrate how the photon/axion mixing inside the source works, we show in Figure~\ref{fig:sourcemix} an example for an AGN modeled by the parameters listed in Table~\ref{tab:3c279}. The only difference is the use of an ALP mass of 1 $\mu$eV instead of the value that appears in that Table, so that we can obtain a critical energy that lies in the GeV energy range; indeed, we get $E_{crit}=0.19$ GeV according to Eq.~(\ref{eq:ecrit}). Note that the main effect just above $E_{crit}$ is an {\it attenuation} of the total expected flux coming from the source.\footnote{Note, however, that the attenuation starts to decrease at higher energies ($>$10 GeV) gradually, the reason being the crucial role of the Cotton-Mouton term, which makes the efficiency of the source mixing to decrease as the energy increases. See Ref.~\cite{masc_axions} for details.}

\subsection{Mixing in the IGMFs}
As already discussed, despite the low magnetic field {\bf B}, the photon/axion oscillation can take place in the IGMFs due to the large distances. However, the strength of the IGMFs is expected to be many orders of magnitude weaker ($\sim$nG) than that of the source and its surroundings ($\sim$G). Consequently, as described by Eq.~(\ref{eq:ecrit}), the energy at which photon/axion oscillation occurs in this case is many orders of magnitude larger than that at which oscillation can occur in the source and its vicinity.  Assuming a mid-value of B$\sim$0.1 nG, and $M_{11}=0.114$ (CAST upper limit), the effect could be observationally detectable by current IACTs ($E_{crit}< 1$~TeV) only if the ALP mass is m$_a<6 \times 10^{-10}$ eV, i.e. we need ultra-light ALPs. For example, we get $E_{crit} = 28.5$ GeV when using m$_a=10^{-10}$ eV, which is the value given in Table \ref{tab:3c279} as our reference one.

 \begin{table}[!ht]
\begin{center}
  \caption{\label{tab:3c279} \small{Parameters used to calculate the total photon/axion conversion in both the source (for the two AGNs considered, 3C~279 and PKS~2155-304) and in the IGM. The values related to 3C~279 were obtained from Ref.~\cite{hartmann}, while those ones for PKS~2155-304 were obtained from  Ref.~\cite{pks2155}. This Table represents our fiducial model.}} 
  \vspace{0.2cm}
    \begin{tabular}{l|l|l|l}
      & Parameter & 3C~279 & PKS~2155-304\\
      \hline
      & B (G) & 1.5 & 0.1\\
      Source & e$_d$ (cm$^{-3}$) & 25 & 160\\
      parameters & {\small L domains (pc)} & 0.003 & 3 $\times$ 10$^{-4}$\\
      & B region (pc) & 0.03 & 0.003\\
      \hline
      & z & 0.536 & 0.117\\
      Intergalactic & e$_{d,int}$ (cm$^{-3}$) & 10$^{-7}$ &  10$^{-7}$\\
      parameters & B$_{int}$ (nG) & 0.1 & 0.1\\
      &  {\small L domains (Mpc)} & 1 & 1\\
      \hline
      ALP & M (GeV) & 1.14 $\times$ 10$^{10}$ & 1.14 $\times$ 10$^{10}$\\
      parameters & ALP mass (eV) & 10$^{-10}$ & 10$^{-10}$\\
    \end{tabular}
\end{center} 
\end{table}

\begin{figure*}[!ht]  \centering
\includegraphics[width=14cm]{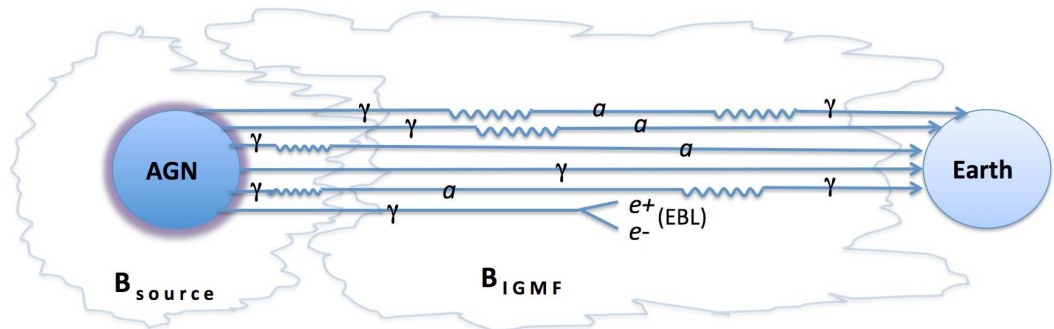}
\caption{\small{Sketch of the formalism here presented, where both mixing inside the source and mixing in the IGMF are considered under the same consistent framework. Photon to axion oscillations (or vice-versa) are represented by a crooked line, while the symbols {\it $\gamma$} and {\it a} mean gamma-ray photons and axions respectively. This diagram collects the main physical scenarios that we might identify inside our formalism. Each of them are squematically represented by a line that goes from the source to the Earth.}}
\label{fig:sketch}
\end{figure*} 

In our model, we assume that the photon beam propagates over N domains of a given length, the modulus of the magnetic field roughly constant in each of them. We take, however, randomly chosen orientations, which in practice is also equivalent to a variation in the strength of the component of the magnetic field involved in the photon/axion mixing\footnote{We refer to Ref.~\cite{masc_axions} for a detailed description of the model and the related equations.}. Moreover, it will be necessary to include the effect of the Extragalactic Background Light (EBL) in the equations as well, its main effect being an additional attenuation of the photon flux, especially at energies above 100 GeV. Indeed, the EBL plays a crucial role to correctly evaluate and understand the importance of the intergalactic mixing. The induced effect can be an {\it attenuation} or an {\it enhancement} of the photon flux at Earth, depending on distance, magnetic field and the EBL model considered (see next Section).
 
 \subsection{Source and intergalactic mixings working together}
In conclusion, AGNs located at cosmological distances will be affected by both mixing in the source  
and in the IGMF. As already mentioned, up to now previous works have focused only in studying the photon/axion mixing either inside the source or in the IGMFs. Instead, for the first time we carried out a detailed study of the mixing in both regimes under the same consistent framework. In order to observe both effects in the gamma-ray band, we need ultra-light ALPs. That is the reason why in our fiducial model, presented in Table~\ref{tab:3c279}, we use a value of m$_a=10^{-10}$ eV, which implies $E_{crit} \sim$ 30 GeV for the IGMF mixing (for B$\sim$0.1~nG) and $E_{crit} \sim$ 1~eV within the source and its vicinity (B$\sim$1~G). Consequently, both effects need to be taken into account. We show in Fig.~\ref{fig:sketch} a diagram that outlines our formalism. Very squematically, the diagram shows the travel of a photon from the source to the Earth in a scenario with ALPs. In the same Figure, we show the main physical cases that one could identify inside our formalism: mixing in both the source and the IGMF, mixing in only one of these environments, the effect of the EBL, etc.

\section{Axion boosts}     \label{sec2}

Since we expect the intergalactic mixing to be more important for larger distances, due to the more prominent role of the EBL, we chose two distant astrophysical sources as our benchmark AGNs: the radio quasar 3C~279 (z=0.536) and the BL Lac PKS~2155-304 at z=0.117. We summarize in Table \ref{tab:3c279} the parameters we have considered in order to calculate the total photon/axion conversion in both the source and in the intergalactic medium. As for the EBL model, we chose the Primack \cite{primack05} and Kneiske best-fit \cite{kneiske04} ones. They represent respectively one of the most transparent and one of the most opaque models for gamma-rays, but still within the limits imposed by the observations.

\begin{figure}[!h]
\centering
  \begin{minipage}[b]{0.48\textwidth}
    \centering
    \includegraphics[width=8cm]{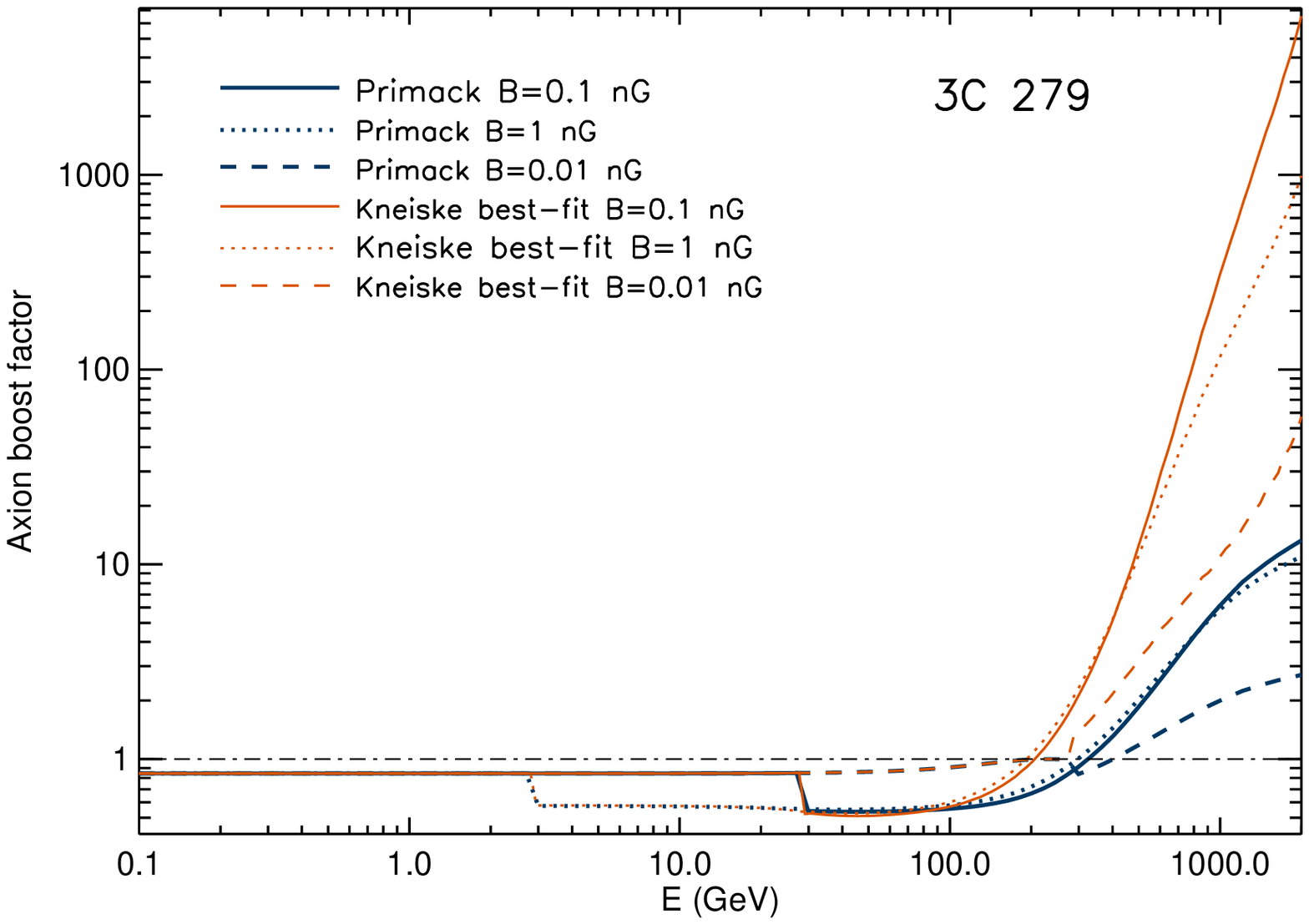}
  \end{minipage}\\
  \begin{minipage}[b]{0.48\textwidth}
    \centering
    \includegraphics[width=8cm]{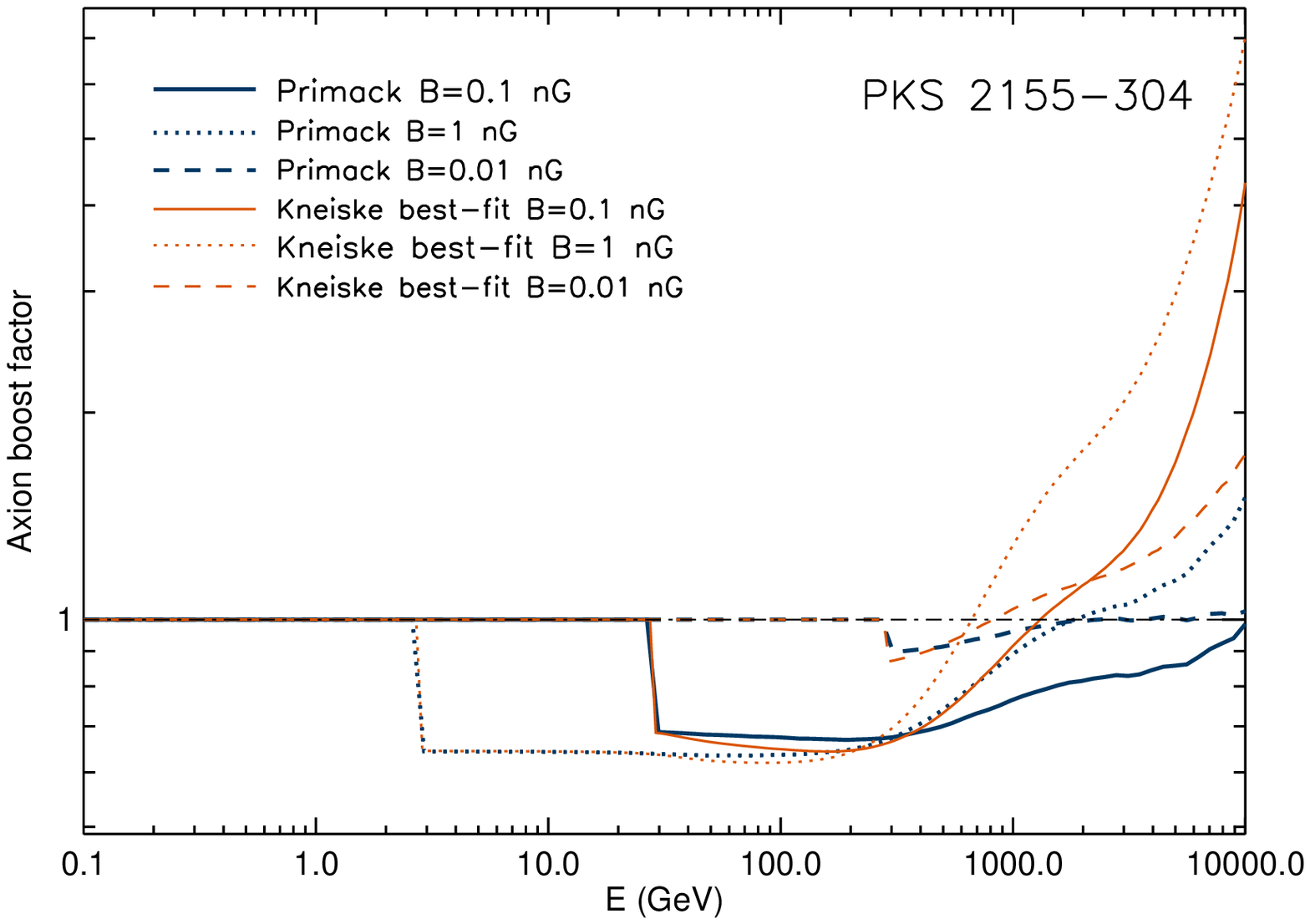}
  \end{minipage}
\caption{\small{Boost in intensity due to ALPs for the Kneiske best-fit (red light lines) and Primack (blue dark lines) EBL models, and for different values of the magnetic field. We used those parameters presented in Table~\ref{tab:3c279} for 3C~279 (z$=$0.536) and PKS~2155-304 (z$=$0.117).}}
\label{fig:boosts}
\end{figure}

\begin{figure*}[!ht]  \centering
\includegraphics[width=14.4cm]{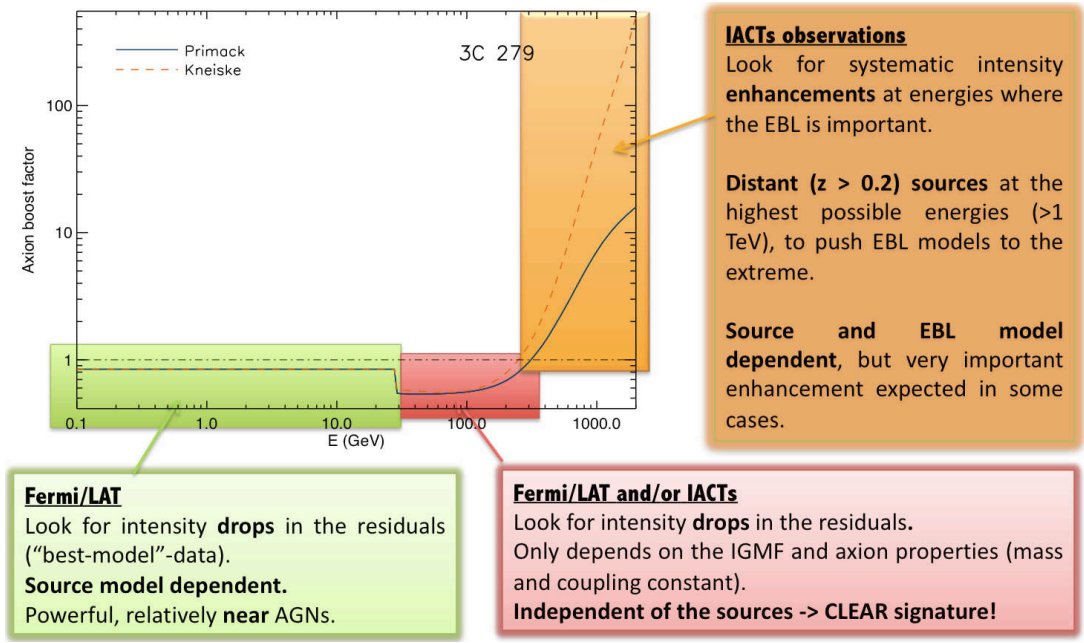}
\caption{\small{Summary of the best observational strategy to look for ALPs with Fermi and IACTs. See Ref.~\cite{masc_axions} for a deeper discussion.}}
\label{fig:obs}
\end{figure*} 

In order to quantitatively study the effect of ALPs on the total photon intensity, we plot in Figure~\ref{fig:boosts} the difference between the predicted arriving photon intensity without including ALPs and that one obtained when including the photon/axion oscillations (called here the {\it axion boost factor}). This was done for our fiducial model (Table~\ref{tab:3c279}) and for the two EBL models considered. The plots show differences in the axion boost factors obtained for 3C~279 and PKS~2155-304 due mostly to the redshift difference. The inferred critical energies for the mixing in the source are $E_{crit}=4.6$ eV for 3C~279 and $E_{crit}=69$ eV for PKS 2155-304, while for the mixing in the intergalactic medium we obtain $E_{crit}=28.5$ GeV (the same for both objects). For B$=$0.1~nG, and in the case of 3C~279, the axion boost is an attenuation of about 16\% below the critical energy (due to mixing inside the source). Above this critical energy and below 200-300 GeV, where the EBL attenuation is still small, there is an extra attenuation of about 30\% (mixing in the IGMF). Above 200-300 GeV the axion boost reaches very high values: at 1 TeV, a factor of $\sim$7 for the Primack EBL model and $\sim$340 for the Kneiske best-fit model. We find that the more attenuating the EBL model considered, the more relevant the effect of photon/axion oscillations in the IGMF, since any ALP to photon reconversion might substantially enhance the intensity arriving at Earth. In the case of PKS~2155-304, the situation is different from that of 3C~279 due to the very low photon-attenuation at the source and, mostly, due to the smaller source distance. Furthermore, a very interesting result is found when varying the modulus of the intergalactic magnetic field. Higher ${\bf B}$ values do not necessarily translate into higher photon flux enhancements. There is always a ${\bf B}$ value that maximizes the axion boost factors; this value is sensitive to the source distance, the considered energy and the EBL adopted model (see Ref.~\cite{masc_axions} for a detailed discussion on this issue).

\section{Detection prospects for Fermi and IACTs}     \label{sec3}

If we accurately knew the intrinsic spectrum of the sources and/or the density of the EBL, we should be able to observationally detect axion signatures or to exclude some portions of the parameter space.  We lack this knowledge, so detection is challenging but we believe that still possible. The combination of the Fermi/LAT instrument and the IACTs, which cover 6 decades in energy (from 20 MeV to 20 TeV) is very well suited to study the photon/axion mixing effect. Nevertheless, and before assuming an scenario with axions to interpret the observational data, we should try to describe the observational data (preferably several AGNs located at different redshifts, as well as the same AGN undergoing different flaring states, from radio to TeV) with ``conventional'' theoretical models for the AGN emission and for the EBL. If  these ``conventional'' models for the source emission and EBL fail (i.e. if we have important residuals for the best-fit model), then the axion scenario should be explored. Fig.~\ref{fig:obs} summarizes what could be a good observational strategy to look for ALPs with Fermi and IACTs.

\section{Are we detecting axions already?}     \label{sec4}

\noindent Recent gamma observations might already pose substantial challenges to the conventional models to explain the observed source spectra and/or EBL density, e.g.:

\begin{itemize}
\item The VERITAS Collaboration recently claimed a detection above 0.1 TeV coming from 3C66A  (z$=$0.444). The EBL-corrected spectrum seems to be harder than 1.5 \cite{acciari09}. 
\item TeV photons coming from 3C 66A? (see Refs. \cite{aliu09,Nesphor1998, Stepanyan2002}). If so, difficult to explain with conventional EBL models and physics. 
\item The lower limit on the EBL at 3.6 $\mu$m was recently revised upwards by a factor $\sim$2, suggesting a more opaque universe \cite{Levenson2008}. 
\item Some sources at z $=$ 0.1-0.2 seem to have harder intrinsic energy spectra than previously anticipated \cite{Krennrich2008}.  
\end{itemize}

\begin{figure}[!h]
\centering
\includegraphics[height=6.5cm,width=8.5cm]{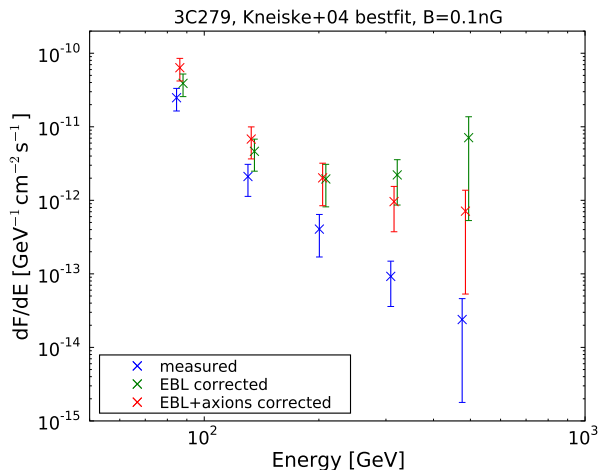}
\caption{\small{Effect of  the existence of ALPs in the intrinsic spectrum of 3C~279, assuming the fiducial model given in Table \ref{tab:3c279} and a Kneiske best-fit model for the EBL. Blue points correspond to observational data taken with the MAGIC telescope \cite{albert08}.}} 
\label{fig:axionsfriends}
\end{figure}

While it is still possible to explain these observations with conventional physics, the axion/photon oscillation would naturally explain these puzzles, since we get more high energy photons than expected as well as a softer intrinsic spectrum in a scenario with ALPs. An example is given in Fig.~\ref{fig:axionsfriends} (from Ref.~\cite{masc&dominguez}), where the effect of the existence of ALPs in the intrinsic spectrum of 3C~279 is shown assuming the fiducial model presented in Table \ref{tab:3c279} and a Kneiske best-fit EBL model.

\section{Conclusions}     \label{sec5}

If axions exist, then they could distort the spectra of astrophysical sources importantly. In addition, if photon/axion mixing occurs in the IGMFs, then a mixing in the source will be at work as well. In particular, for axion masses of the order or 10$^{-10}$ eV, the induced effect is expected to be present in the gamma ray energy range. 

Since photon/axion mixing in both the source and the IGM are expected to be at work over 
several decades in energy, it is clear that a joint effort of Fermi and current IACTs is needed. The Fermi/LAT instrument is expected to play a key role, since it will detect thousands of AGNs (up to z$\sim$5), at energies where the EBL is not important.  By the other hand, IACTs will be specially important at higher energies ($>$300 GeV), where the EBL is present.

\noindent Main caveats: the effect of photon/axion oscillations could be attributed to conventional 
physics in the source and/or propagation of the gamma-rays towards the Earth. However, we believe that detailed observations of AGNs at different redshifts and different flaring states could be successfully used to identify the signature of an effective photon/axion mixing.

\end{document}